\documentstyle[12pt,epsf,epsfig]{article}
\newfont{\thiplo}{msbm10 scaled\magstep 2}
\newfont{\gothic}{eufb10 scaled\magstep 2}
\newfont{\unc}{eurb10} 
\newskip\humongous \humongous=0pt plus 1000pt minus 1000pt
\def\caja{\mathsurround=0pt}\def\eqalign#1{\,\vcenter{\openup1\jot \caja
        \ialign{\strut \hfil$\displaystyle{##}$&$ 
        \displaystyle{{}##}$\hfil\crcr#1\crcr}}\,}
\newif\ifdtup

\def\eqright #1\cr{\noalign{\hfill$\displaystyle{{}#1}$}}
\def\eqleft #1\cr{\noalign{\noindent$\displaystyle{{}#1}$\hfill}}
\def\oldreffmt#1{\rlap{[#1]} \hbox to 2\parindent{}}

\def\figfmt#1{\rlap{Figure {#1}} \hbox to 1in{}}

\def\sectioneq{\def\theequation{\thesection.\arabic{equation}}{\let
\holdsection=\section\def\section{\setcounter{equation}{0}\holdsection}}}%

\newcounter{holdequation}

\def\begineq #1\endeq{$$ \refstepcounter{equation}\eqalign{#1}\eqno
	(\theequation) $$}
\def\contlimit{\,{\hbox{$\longrightarrow$}\kern-1.8em\lower1ex
\hbox{${\scriptstyle (a\rightarrow0)}$}}\,}
\def\centeron#1#2{{\setbox0=\hbox{#1}\setbox1=\hbox{#2}\ifdim
\wd1>\wd0\kern.5\wd1\kern-.5\wd0\fi
\copy0\kern-.5\wd0\kern-.5\wd1\copy1\ifdim\wd0>\wd1
\kern.5\wd0\kern-.5\wd1\fi}}
\def\centerover#1#2{\centeron{#1}{\setbox0=\hbox{#1}\setbox
1=\hbox{#2}\raise\ht0\hbox{\raise\dp1\hbox{\copy1}}}}
\def\centerunder#1#2{\centeron{#1}{\setbox0=\hbox{#1}\setbox
1=\hbox{#2}\lower\dp0\hbox{\lower\ht1\hbox{\copy1}}}}
\def\lsim{\;\centeron{\raise.35ex\hbox{$<$}}{\lower.65ex\hbox
{$\sim$}}\;}
\def\gsim{\;\centeron{\raise.35ex\hbox{$>$}}{\lower.65ex\hbox
{$\sim$}}\;}
\def\super#1{\ifmmode \hbox{\textsuper{#1}}\else\textsuper{#1}\fi}
\def\textsuper#1{\newcount\holdspacefactor\holdspacefactor=\spacefactor
$^{#1}$\spacefactor=\holdspacefactor}
\def\getcite#1,{\advance\citenumber by1
\def\getcitearg{#1}\def\lastarg{@}
\ifnum\citenumber=1
\ref{#1}\let\next=\getcite\else\ifx\getcitearg\lastarg\let\next=\relax
\else ,\ref{#1}\let\next=\getcite\fi\fi\next}
\def\pom{{\rm P\kern -0.53em\llap I\,}}
\def\spom{{\rm P\kern -0.36em\llap \small I\,}}
\def\sspom{{\rm P\kern -0.33em\llap \footnotesize I\,}}

\relax
\def\contlimit{\,{\hbox{$\longrightarrow$}\kern-1.8em\lower1ex
\hbox{${\scriptstyle (a\rightarrow0)}$}}\,}
\def\upon #1/#2 {{\textstyle{#1\over #2}}}
\relax
\renewcommand{\thefootnote}{\fnsymbol{footnote}} 

\def\mainhead#1{\setcounter{equation}{0}\addtocounter{section}{1}
  \vbox{\begin{center}\large\bf #1\end{center}}\nobreak\par}
\sectioneq

\def\til#1{\centeron{\hbox{$#1$}}{\lower 2ex\hbox{$\char'176$}}}
\def\tild#1{\centeron{\hbox{$\,#1$}}{\lower 2.5ex\hbox{$\char'176$}}}
\def\sumtil{\centeron{\hbox{$\displaystyle\sum$}}{\lower
-1.5ex\hbox{$\widetilde{\phantom{xx}}$}}}

\begin{document} 

\begin{titlepage} 

\rightline{\vbox{\halign{&#\hfil\cr
% &ANL-HEP-CP-04-83\cr
&\today\cr}}} 
\vspace{0.25in} 

\begin{center} 
  
{\large\bf A Higgs-Like Scalar With Mixed-Parity and Standard-Model-Like Couplings
Could be Further Evidence for Underlying Massless SU(5) Unification. }

\medskip

Alan R. White\footnote{arw@anl.gov }

\vskip 0.6cm

\centerline{Argonne National Laboratory}
\centerline{9700 South Cass, Il 60439, USA.}
\vspace{0.5cm}

\end{center}\begin{abstract} 

The bound-state S-Matrix of {\bf QUD}  \{SU(5) gauge theory with massless 
left-handed $ {\bf 5 \oplus 15 \oplus 40 \oplus 45^*}$
fermions\} might underly the success of the Standard Model. The dynamical role of the QUD top quark leads to two Higgs-like scalar resonances, the $\eta_6$ - which reproduces Standard Model top events, and the $\eta_3$ - which could have a mass $\sim$ 125 GeV.
The participation of the weak interaction in the dynamics
implies the resonances are not parity eigenstates and so should have pseudoscalar and scalar electroweak couplings. Also, a tree-unitarity condition could hold 
- in part because of the closeness of the electroweak physical amplitudes to perturbation theory, and - in part because of the intimate relation with regge behavior. This would imply that the combined physical couplings of the two resonances are comparable to those of the Standard Model Higgs boson. Unlike the isolated Standard Model, 
the underlying unification of {\bf QUD} implies there will be no LHC ``nightmare scenario'' and that, instead, a broad, extensive, experimental program should, eventually, be implemented.
\end{abstract} 

\renewcommand{\thefootnote}{\arabic{footnote}} \end{titlepage} 

\mainhead{1. INTRODUCTION}

It appears that both CMS\cite{CMS} and ATLAS\cite{ATL} have discovered a Higgs-like neutral boson with physical couplings that are close to those expected if it is associated with electroweak symmetry breaking. 
Although, these couplings have not yet been shown to have the equality that would provide clear evidence of a connection to a scalar condensate that is the origin of general particle masses. From Weinberg's original paper\cite{swe} onwards, it has been widely anticipated that the existence of a ``Higgs boson'' would be correlated with the existence of a neutral scalar field whose condensate, not only provides 
mass generation for the electroweak bosons, but also gives masses to all elementary fermions. In fact, as is now well-known, the quantum theory of a simple scalar field has so many problems that it is very unlikely that such a theory could provide a condensate that generates 
masses\footnote{Quite apart from the profound philosophical concern that this 
would be a worryingly simplistic ``ether-like'' explanation of the origin of mass.}. This has been a major motivation for the various extensions of the Standard Model, particularly via supersymmetry that have a more elaborate Higgs sector which is directly linked to other sectors. However, the LHC has, so far, failed to provide evidence for any of the proposed extensions, supersymmetric or otherwise. 
Indeed, if the LHC results continue to suggest that something close to the 
``stand-alone'' Standard Model is present in nature, but without the condensate mass generation established; rather than producing the ``nightmare scenario''\footnote{
The LHC discovery of just a Standard Model Higgs boson, and nothing else.},
% is commonly referred to as the ``Nightmare Scenario''.}, 
this would increase the likelihood that the bound-state
S-Matrix of a massless SU(5) gauge theory\cite{kw} that I call QUD\footnote{{\bf QUD} $\leftrightarrow$ Quantum Uno/Unification/Unitary/Underlying Dynamics $\equiv$ SU(5) gauge theory with
massless left-handed $ 5 \oplus 15 \oplus 40 \oplus 45^*$ fermions.} is providing the underpinning, as I have proposed\cite{arw10}-\cite{arw12}. In which case, as I discuss in the last Section, there could eventually be a very exciting and broad experimental program at 
the LHC.

After a long search for a  gauge theory in which reggeon interactions could reproduce the
unitary Critical Pomeron\cite{cri},
I arrived, uniquely, at QUD.
It was a stunning realization that the massless fermion anomaly dynamics of the QUD S-Matrix might also provide a complete, economic, and
aesthetically appealing, unification of the forces and particles that comprise 
the full Standard Model. The only drawback, which is social and philosophical rather than scientific, is that a radical departure from the current theory paradigm has to be 
accepted\footnote{The QUD S-Matrix is an isolated and self-contained description of particle physics - without the off-shell amplitudes usually focussed on in the conventional quantum field theory paradigm.}. Moreover, even though the path I followed is littered with wrong turns and incomplete arguments, there is no ambiguity in the outcome. QUD is a unique theory which can not be added to or modified
in any way. If it is wrong, there is no 
fallback position. 

Unfortunately, the physical S-Matrix can presently only be accessed\cite{arw10,arwdm} via the 
elaborate, but still very underdeveloped, multi-regge reggeon diagram formalism\footnote{Currently, this powerful formalism is (regrettably) well
outside the litany of tools used by most mainstream theorists. In addition, I have
used my extensive study of both the formalism and it's background to maximally push it to the frontier of it's applicability.} that I have utilised. Consequently,  
there is still considerable ambiguity - as the present paper illustrates well - in determining detailed properties. The major new physics is provided by a higher mass, strongly interacting, sector that (as I have constantly lamented) can not be searched for directly
in the current high luminosity LHC program. Trying to extract short-distance properties that might be seen distinctly in the latest LHC results is, at a minimum, a risky exercise.

Within QUD, all particle masses and all the symmetry breaking are S-Matrix properties that (in contrast to the Standard Model) do not appear in the field theory lagrangian.
Instead, these properties appear only as the asymptotic states, and associated scattering amplitudes, are formed via the underlying  
infra-red anomaly dominated dynamics. In a previous paper, I have argued that there are two Higgs-like boson resonances,
the $\eta_6$ and the $\eta_3$, that (because they are expected to mix significantly) should both be closely inter-related with the electroweak symmetry breaking.
In this paper, I will describe likely parity properties of these resonances that may be essential if 
a potential explanation for the latest LHC results is to be provided.

The $\eta_6$ is predominantly the ``sextet $\eta$'', a U(1) neutral pseudoscalar Goldstone boson that is a remnant of the color sextet electroweak symmetry breaking that, at first sight, looks analagous to the Standard Model Higgs boson. Mixing with the pomeron sector breaks the U(1) chiral symmetry, not only giving the $\eta_6$ a mass, but also providing the mixing with the $\eta_3$. The $\eta_3$ is predominantly a triplet quark $t_R\bar{t}_L$ pair, where $t_R$ is a right-handed QUD top quark that (crucially for the existence of the $\eta_3$) is very different\cite{arw12} from the Standard Model top quark. It is a central argument of this paper that,
because the $t_L\bar{t}_R$ component is suppressed relative to the $t_R\bar{t}_L$ component, the $\eta_3$ is not a parity eigenstate, and so, because of the mixing, neither is the $\eta_6$. 

Within the QUD S-Matrix, the $\eta_6$ is responsible for events identified as Standard Model top quark/antiquark production and so it must have a ``sextet scale'' mass $\sim$ 330 GeV. The mixed parity property should provide 
interference with the positive parity, charge asymmetric, background produced by the initial state of the Tevatron that would result in an asymmetry of the kind observed\cite{CDFt}.
Via both the parity breaking effects of the electroweak interaction and 
the mixing with the $\eta_6$, the $\eta_3$  should have a mass between the triplet and sextet mass scales that, obviously, could be $\sim$ 125 GeV. If the $\eta_3$ is actually
the Higgs-like boson discovered at the LHC then increasingly accurate parity
measurements should observe the mixed parity. This would then add to a number of phenomena that I have previously identified\cite{arwdm,arw12} as evidence for the existence of the sextet quark sector and the underlying presence of QUD.

In \cite{arw12}, I described the $\eta_3$ and $\eta_6$ resonances as pseudoscalars.
Unfortunately, I did not recognize that the electroweak coupling of the $t$ to
an exotic quark sector (that has either very high mass bound states or none at all) should not only eliminate all $t$ bound-states with other quarks but should also suppress the weakly interacting $t_L\bar{t}_R$ component of the $\eta_3$ and the $\eta_6$. This is why these resonances are not simple pseudoscalars. In fact, the existence of scalar couplings, when combined with the closeness of the electroweak reggeon diagram amplitudes to perturbation theory, makes it more likely 
that a ``tree-unitarity'' constraint holds. The intimate relation of this constraint to the regge behavior of the physical S-Matrix is an additional, strong, reason why it should hold. It is of crucial relevance because it requires the combined physical electroweak couplings of the two resonances to be comparable to those of the Standard Model Higgs boson. This would be the case, even though the character of the QUD electroweak symmetry breaking is very different from the Standard Model.

The full reggeon diagram construction of the high-energy QUD S-Matrix is a
giant task which I hope to discuss further in a future paper. So far I have 
provided only an outline\cite{arw10,arwdm} for how the Standard Model interactions and low mass spectrum emerge.
In addition to a variety of points of principle that need to be
better resolved, there is a wide array of anomaly vertices that must be explicitly derived and for which summation techniques must be devised. (This is essential if the schematic diagrammatic descriptions that are the essence of my current arguments are to be more concretely developed.) Nevertheless, if it becomes evident from experimental evidence that the QUD S-Matrix could indeed be the much sought after origin of the Standard Model, the reggeon diagram construction will surely be rapidly pursued, with great intensity, by the community at large. While, at present, only limited conclusions can be drawn with respect to the mass spectrum, cross-sections etc., the outline construction is sufficient to allow at least a partial discussion of why tree-unitarity could hold.

\mainhead{2. THE UNNECESSARY HIGGS CONDENSATE}

To generate massive vector bosons it is sufficient that the, initially massless,
vectors mix with massless scalar or pseudoscalar Goldstone bosons that have the right quantum numbers to provide the longitudinal components of the, ultimately massive, vectors. A priori, there is no necessity for the symmetry breaking that is involved to be due to a field condensate. In QUD, instead, it is chiral symmetry breaking by zero momentum chirality transitions, in reggeon anomaly vertices, which produces ``sextet pion'' pseudoscalar Goldstones as anomaly poles. For the resulting anomaly-based mass generation to work, it is crucial that the couplings of the electroweak bosons are left-handed  (which is not, of course, the case for the Standard Model Higgs mechanism).

There is no direct connection between the electroweak vector boson masses and the dynamical masses of the other physical particles, all of which are S-Matrix bound-states. The anomaly vertex chirality transitions appear only in very special reggeon interactions, which include those that generate the left-handed vector boson masses, but otherwise are only indirectly related to general bound-state masses. They do not have the ubiquitous behavior of the field condensate that appears in the Standard Model lagrangian. The chirality transitions are, nevertheless, directly responsible for symmetry breaking in the physical S-Matrix and are a central element of the formation of a massive spectrum.

\mainhead{3. TREE UNITARITY AND REGGEIZATION}

Not very long after the discovery of the
Weinberg-Salam theory it was realized that if a non-abelian massive vector boson 
theory is considered alone, then unitarity bounds for the perturbative amplitudes (tree-unitarity) necessitate\cite{lls} the intermediate state production of, at least, an additional scalar particle. At high energy, this production is needed to cancel the unitarity violating intermediate state vector boson production. If the vector boson symmetry group is SU(2) (as in the Standard Model) and if there is just one neutral scalar then, 
remarkably perhaps, it's couplings must be exactly those of the Higgs boson in the Weinberg-Salam model. In effect, the gauge theory formulation can be by-passed as the origin of the Higgs boson by simply appealing to the unitarity of the S-Matrix
(which is, of course, gauge-invariant). If there are more scalars, they must combine, at high energy, to reproduce the contributions of the single scalar.

The contributions to the longitudinal vector boson scattering amplitude, in the zero charge channel, from the three interactions described by the non-abelian vector interaction are shown
in Figs.~1(a), (b) and (c). 
The scalar production fourth term, shown in Fig.1(d), not only cancels the unitarity violating high energy and angular dependence of the first three terms (as is shown explicitly in \cite{mt}) but, as illustrated, leaves on-shell vector exchange as the remaining high-energy amplitude.

That on-shell vector exchange provides the lowest-order high-energy amplitude is a fundamental result since it is the starting point for the perturbative reggeization that leads to the description of high-energy amplitudes via reggeon diagrams. Indeed, while the Weinberg-Salam theory was
first discovered via the search for a gauge invariant, renormalizable, 
massive vector boson field theory, it subsequently became clear that it could be formulated directly\cite{lnl}-\cite{arw00} via the reggeization and unitarity properties of the S-Matrix. 
The renormalizability of the gauge theory is closely related to the existence of 
t-channel dispersion relations that provide\cite{arw00} the underpinning of regge behavior. 
Indeed, that on-shell vector exchange describes elestic scattering can be derived directly\cite{lnl,arw93} from a t-channel dispersion relation, without any recourse to the summation of the diagrams of Fig.~1.
The most significant consequence, which is not as well-known as it surely should be, is that multiparticle t-channel unitarity is explicitly satisfied\cite{arw93,arw00} by the full set of reggeon diagrams describing the multi-regge behavior of all multiparticle amplitudes.
Since my construction of the QUD high-energy S-Matrix is entirely based on reggeon diagrams involving both gauge bosons and fermions, the cancelation implied by tree-unitarity is a basic ingredient.
\begin{center}
\epsfxsize=5in \epsffile{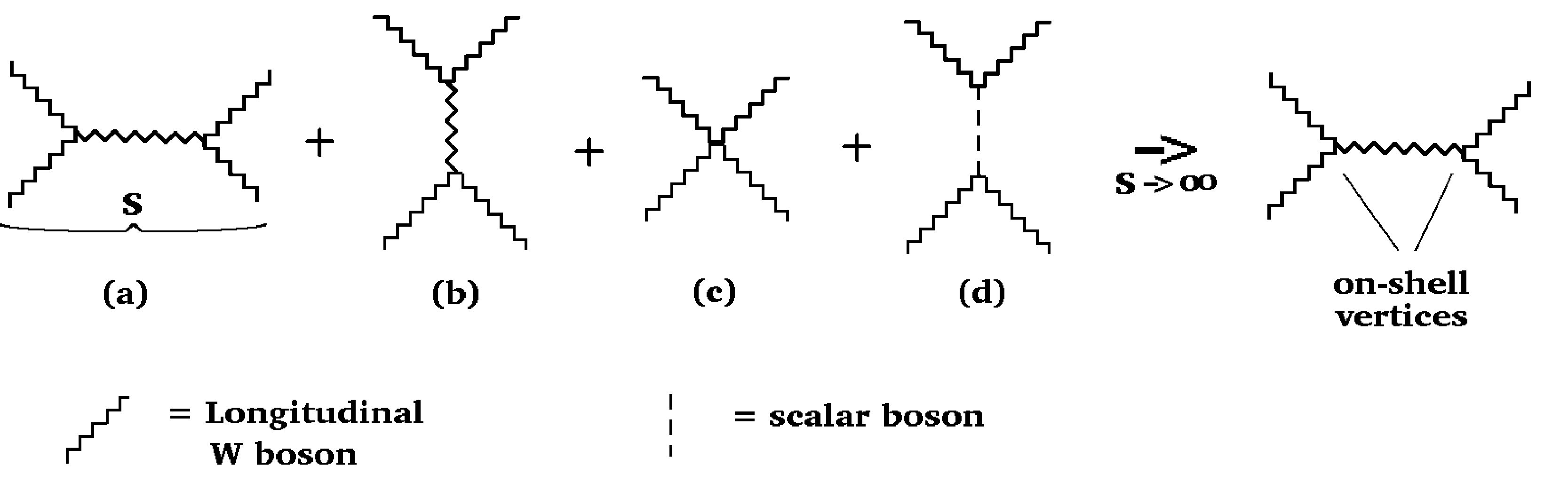}  

Figure 1 Tree Scattering Amplitudes (a) vector boson exchange (b) vector boson production (c) four vector interaction (d) scalar boson production

\end{center}

While the perturbative tree-unitarity bounds are satisfied by the electroweak sector of the Standard Model, they are not expected to be applicable in general non-perturbative dynamical theories, such as technicolor models. Similarly, at first sight, it might be expected that, as the ``non-perturbative'' dynamics of the QUD S-Matrix emerges, the underlying presence of perturbative tree-unitarity would be obscured. However, as I have emphasized in previous papers and will illustrate specifically in the following Sections, as they are built up, the 
physical reggeon diagrams remain very close to the perturbative diagrams, particularly in the electroweak sector. The full gauge symmetry is built up in stages and, in the first stage, (essentially)
perturbative amplitudes are accompanied by wee parton scattering processes involving anomaly couplings. The anomaly generation of vector boson
masses is already in place at this stage and, as the full symmetry is restored, color factors
imply that the sextet sector dominates. Consequently, the $\eta_6$, and the $\eta_3$ (because of it's sextet $\eta$ component), are the only states able to provide the  cancelation needed for regge behavior to emerge via underlying tree-unitarity. They are directly produced in the electroweak boson interaction in parallel with Standard Model Higgs boson production and so, as I will discuss further in Section 8, I anticipate that the regge behavior of physical QUD S-Matrix amplitudes requires that couplings similar to those of the Standard Model Higgs boson should emerge.

\mainhead{4. QUD S-MATRIX INTERACTIONS AND STATES}

QUD is the only\cite{kw} unitary gauge theory that is, anomaly free and asymptotically free, contains both the electroweak and QCD interactions, and contains the sextet sector required for electroweak symmetry breaking.
I have argued that all these features are needed to obtain a unitary high-energy S-Matrix via the Critical Pomeron. 

Under $SU(3)\otimes SU(2)\otimes U(1)$ the contributing representations break down as
\begin{center}
\openup0.75\jot
{\footnotesize ~~~${\bf 5=[3,1,-\frac{1}{3}]^{3}
+[1,2,\frac{1}{2}]^{ 2}~,~~~~~ 15=[1,3,1]+
[3,2,\frac{1}{6}]^{1}+ \{6,1,-\frac{2}{3}\}^{\#}~,}$}
{\footnotesize $ {\bf 40=[1,2,-\frac{3}{2}]^{3}
+[3,2,\frac{1}{6}]^{2}+
[3^*,1,-\frac{2}{3}]~+~[3^*,3,-\frac{2}{3}]+
\{6^*,2,\frac{1}{6}\}^{\#}+[8,1,1]~,}$}\\
{\footnotesize ${\bf
 45^*=[1,2,-\frac{1}{2}]^{1}+[3^*,1,\frac{1}{3}]
+[3^*,3,\frac{1}{3}]+[3,1,-\frac{4}{3}]
+[3,2,\frac{7}{6}]^{3}+
\{6,1,\frac{1}{3}\}^{\#} +[8,2,-\frac{1}{2}]}$}
\end{center}
The sextet quarks \{...\}$^{\#}$ have just the right quantum numbers for sextet ``pions'' to provide the longitudinal components of the electroweak vector bosons. Remarkably, both
the triplet quark and lepton sectors, are amazingly close to the Standard Model. 

Under the same decomposition, the gauge bosons give 

\vspace{0.1in}
\centerline{\footnotesize $ {\bf 24=[1,1,0] ~+ ~[1,3,0]
~+~[3^*,2,\frac{5}{6}] ~+~
[3,2,-\frac{5}{6}]~+~[8,1,0]}$}

\noindent and so they can directly transform triplet quarks, but not sextet quarks, to leptons. 
This is why massive triplet quarks would be generated (by self-energy corrections) 
as a component of a (Standard Model?) effective lagrangian if 
the bound-state massive physical leptons could be introduced by ``integrating out'' the elementary massless leptons. In first approximation at least, the sextet quarks would remain massless.

As can be seen from the above decomposition, QUD is real (vector-like) with respect to $SU(3) \otimes U(1)_{em}$ and it contains three generations 
of quarks with charges (2/3,~-1/3) - denoted by superscripts {\bf 1,2,3}.
It also contains higher charge quarks and antiquarks that are crucial for the 
``top quark dynamics'', partially discussed in \cite{arw12} and \cite{arwdm},
and discussed further in later Sections. The ``QUD top'' is identified as 
the charge 2/3 quark in the $[3,2,\frac{7}{6}]$ doublet and so it couples, via the electroweak interaction, to the ``exotic'' higher charge sector.
As elaborated in \cite{arw10},
the octet quarks play a fundamental role in producing an infinite transverse
momentum anomaly contribution that is responsible for the emergence of bound-state lepton and hadron generations with Standard Model quantum numbers. 

A priori, since QUD is a massless field theory, we might expect that
any ``non-perturbative'' dynamics would be prohibitively
difficult to investigate. Indeed, it is very fortunate that multi-regge theory provides access\cite{arw10} to the bound-state S-Matrix. Special properties of QUD lead to 
``universal wee partons'', arising from infinite momentum infra-red (transverse momenta) divergences, which play a vacuum role in the reggeon diagram construction of bound-state amplitudes. As we have described in previous papers, the exponentiation of reggeization divergences, coupled to the infra-red and ultra-violet 
behavior of the effective vertices provided by massless fermion loops, eliminates the vast majority of reggeon diagrams. Only the scaling infra-red divergence due to very special ``anomalous'' (signature $\neq$ color parity) wee gauge boson configurations remains and this is factored off to obtain physical amplitudes. The bound-states are produced, either wholly or in part, by ``anomaly poles'' which are a consequence of the infra-red anomalies of massless fermion loop reggeon effective vertices coupled to this divergence. 

The reggeon diagram infra-red analysis begins with the gauge symmetry broken by
both reggeon masses and a transverse momentum cut-off. The order of removal of the masses and the cut-off is crucial\cite{arw10}, but the reggeon diagram components of the final amplitudes are, after an
averaging over the direction of the symmetry breaking, singlets with respect to the global SU(5) symmetry. The most straightforward surviving interactions are, in first
approximation, SU(5) singlet reggeon states that contain a transverse momentum carrying vector boson exchange accompanied by anomalous wee gauge bosons. They are 
\begin{itemize}
\item{{\bf The Even Signature Pomeron -} {\it an SU(3) color sub-group gauge boson accompanied by odd signature anomalous wee gauge boson
configurations.}}
\item{{\bf The Odd Signature Photon -} {\it an SU(3) color singlet, abelian, gauge boson accompanied by even signature anomalous wee gauge boson
configurations.}}
\end{itemize}
The analagous exchanges of gauge bosons with left-handed fermion couplings are eliminated altogether by interactions with the anomalous wee gauge bosons. There is, however, an additional interaction that is the central concern of this paper. i.e.
\begin{itemize}
\item{{\bf The Electroweak Interaction -} {\it
a color singlet left-handed gauge boson, that aquires both a mass and a flavor
by mixing with massless sextet pions, accompanied by 
odd signature anomalous wee gauge boson configurations.}}
\end{itemize}

To illustrate the physical effect of the ``wee parton vacuum'', we show, in Fig.~2 a lowest-order amplitude describing electron scattering via photon exchange.
Because of it's relative simplicity, we can use this process to illustrate the complexity of the wee parton structure. The resulting physical amplitude is very close to the familiar perturbative
amplitude (as it has to be, of course) apart from the, obviously essential, renormalization via anomaly factors
that implies the coupling is not the very small QUD coupling.
\begin{center}
\parbox{0.4in}{(a)
\vspace{1in}}
\parbox{5.5in}{ \epsfxsize=5.4in \epsffile{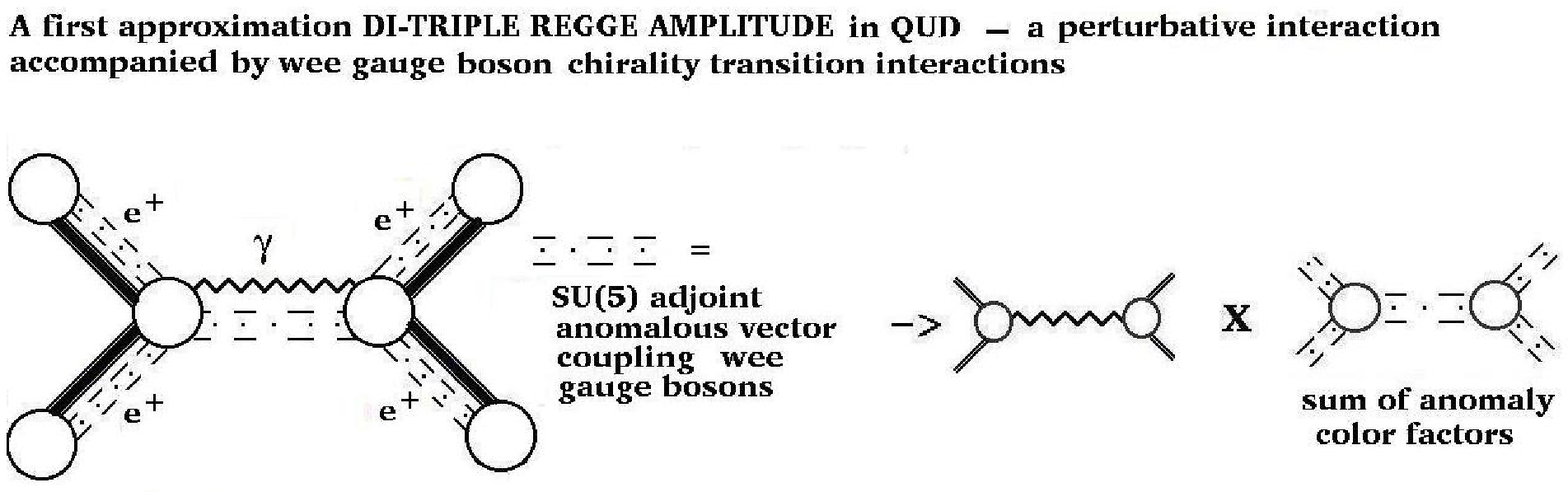}}

\parbox{0.3in}{$~$ \newline (b)
\vspace{1.5in}}
\parbox{5.6in}{ \epsfxsize=5.5in \epsffile{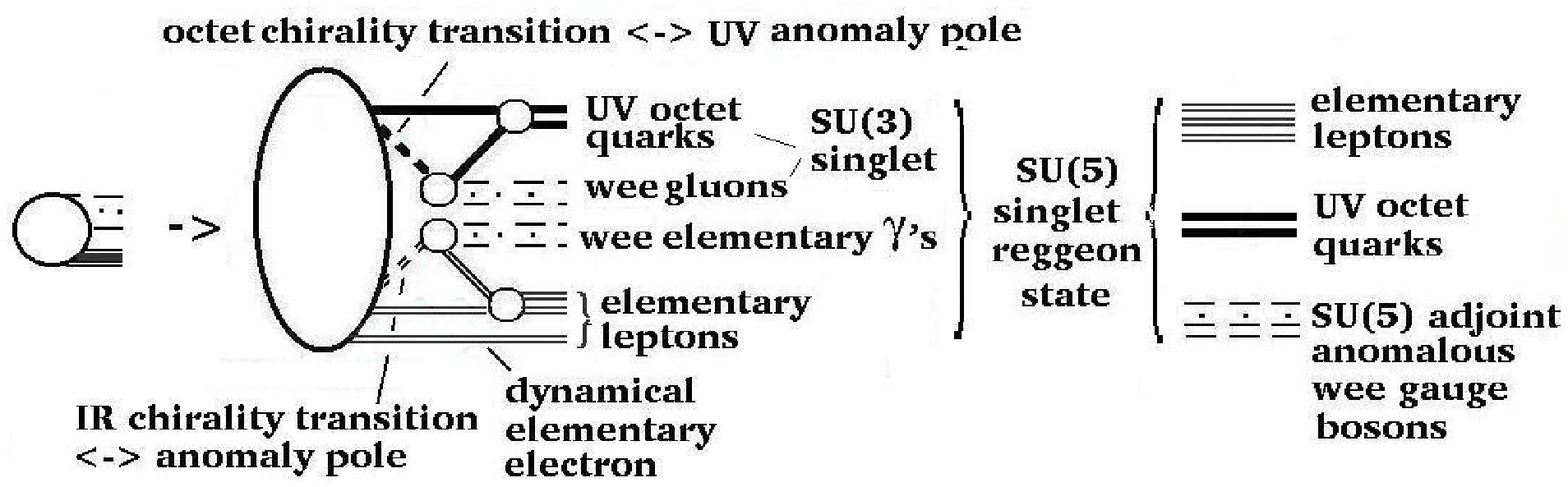}}

\parbox{0.3in}{(c)
\vspace{2.5in}}
\parbox{5.6in}{ \epsfxsize=5.5in \epsffile{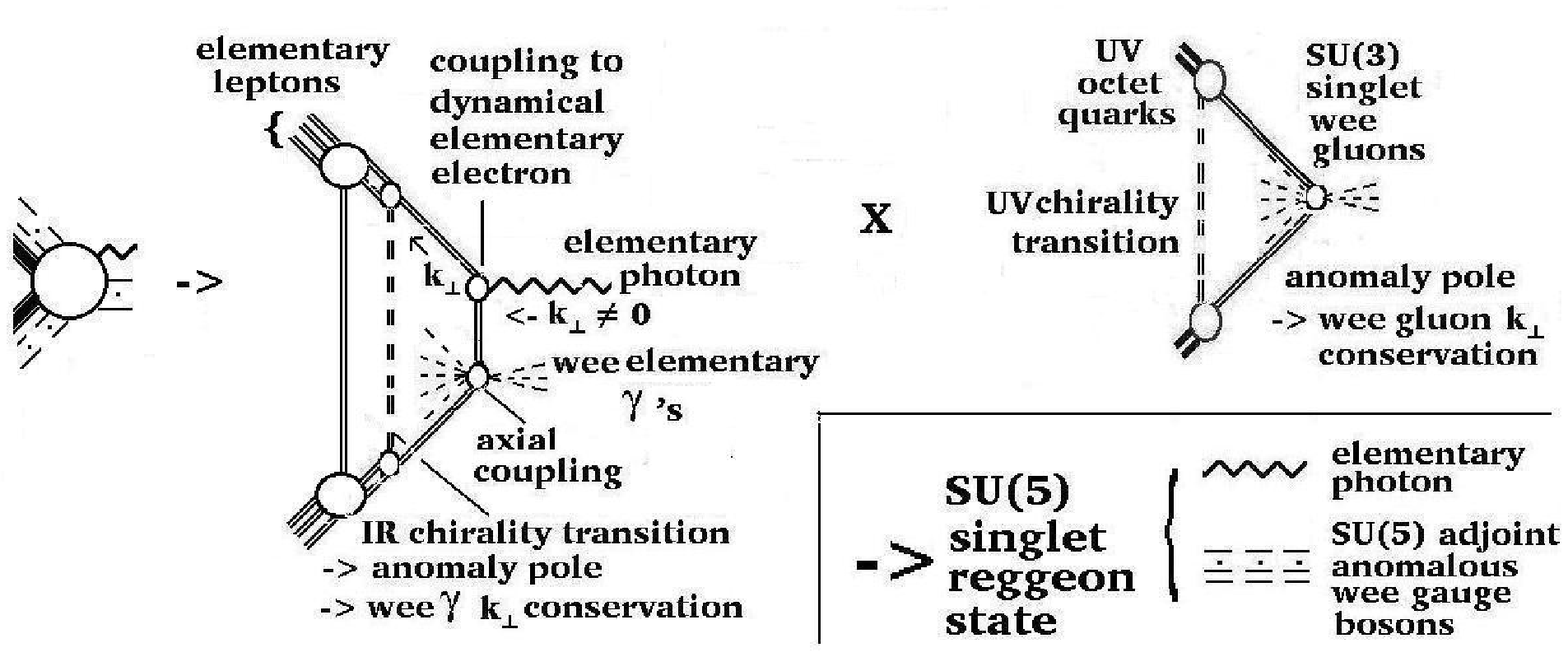}}

Figure 2 Electron Scattering Via Photon Exchange in QUD (a) The Amplitude (b)
The Physical Electron Reggeon State (c) The Physical Photon
\end{center} 

As shown in Fig.~2(a), the basic wee parton component contains general configurations of ``anomalous''  wee gauge bosons, each of which must have vector fermion couplings, and which collectively have an SU(5) adjoint projection. 
All the wee gauge bosons carry zero transverse momenta. 
In general bound-states the fermion content, like the ``elementary photon'', forms an SU(5) adjoint projection that combines with the wee gauge bosons to make an SU(5) singlet projection (with global SU(5) invariance obtained by summation over projections).

Fig.~2(b) illustrates how the fermion content of the electron state is built up.
As shown, there are three elementary leptons, of which two are produced by an anomaly pole and one is dynamical. In lowest-order the dynamical elementary lepton is responsible for the interaction. (The larger mass muon and tau leptons
are more dependent on higher-order interactions and so are not so simple.)
 An infinite transverse momentum pair of color octet quarks 
is present as an anomaly contribution (equivalent 
to an ultra-violet anomaly pole). They do not participate in dynamical interactions but produce an SU(5) invariant reggeon state and, effectively, determine the electroweak interaction quantum numbers of the physical states. As shown, although the electron reggeon state is SU(5) invariant, it is connected to the physical electron bound state, that is produced by a combination of the dynamical elementary electron and anomaly poles, via zero momentum chirality transitions that break the global SU(5) symmetry in the physical bound state. 

The vertex interaction illustrated in Fig.~2(c) is simpler, and more obviously 
does what is required, than the process that has appeared in my previous papers.
A chirality transition of one of the additional elementary leptons in the scattering state produces an anomaly pole that couples the wee
photons in the physical electron and physical photon.The elementary photon couples to the
dynamical electron but, as shown, the transverse momentum flow has to avoid
the triangle of propagators responsible for the infra-red anomaly pole. (Two of the propagators must carry the same light-like momentum while the third, involving the chirality transition, carries zero momentum).  
The infinite transverse momentum anomaly pole produced by the octet quarks 
again couples the SU(3)
color singlet component of the anomalous wee gauge bosons. 

As is apparent, even in lowest-order, a relatively complicated wee parton interaction, within the ``universal wee parton vacuum'' provides a background to the elementary photon exchange process. In higher-orders ever more complicated
background processes will obviously contribute.
As we have discussed\cite{arw10,arwdm}, higher-order contributions contain many anomaly vertices involving the wee parton component of the scattering states, as illustrated, for example, in Fig.~3. 
The additional interactions
are relatively insignificant for the photon and for the electroweak vector bosons do little more than produce a mass, as we discuss next. For the pomeron, both the transverse momentum carrying gauge boson and the anomalous wee gauge boson component carry non-abelian SU(3) color. Consequently, as discussed at length in other papers,
the pomeron couples strongly, via anomaly vertices, to hadron bound-states and has an anomaly-dominated triple pomeron vertex. 

In general, the higher-order diagrams are
essential to build up the complete set of massive physical states, as well as all 
scattering amplitudes.
Since the QUD perturbation expansion has no renormalons, it is possible that summation procedures could be developed to obtain all-orders results for various
classes of diagram sums.
\begin{center}
\epsfxsize=5.3in\epsfbox{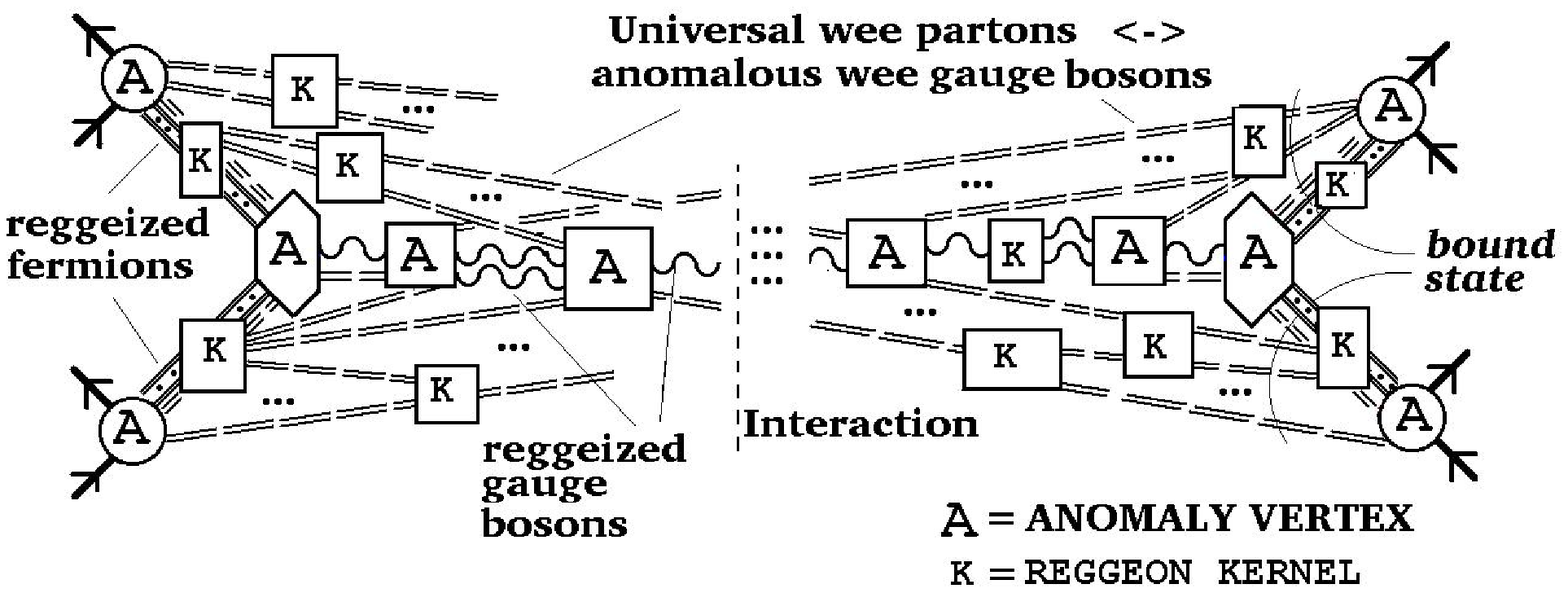}

Figure 3 A Higher-Order Di-Triple Regge Amplitude in QUD
\end{center}

\mainhead{5. VECTOR BOSON MASS GENERATION}

For the left-handed electroweak bosons, the anomaly coupling 
to bound-states, analagous to Fig.~2(c), allows an exchanged boson to mix directly with pseudoscalar chiral Goldstone bosons. This will happen as part of the infra-red analysis, once this interaction is introduced - following the first SU(2)$_C$ color symmetry restoration. The result is vector boson  mass generation of the form
illustrated in Fig.~4.
\begin{center}
\epsfxsize=4.5in \epsffile{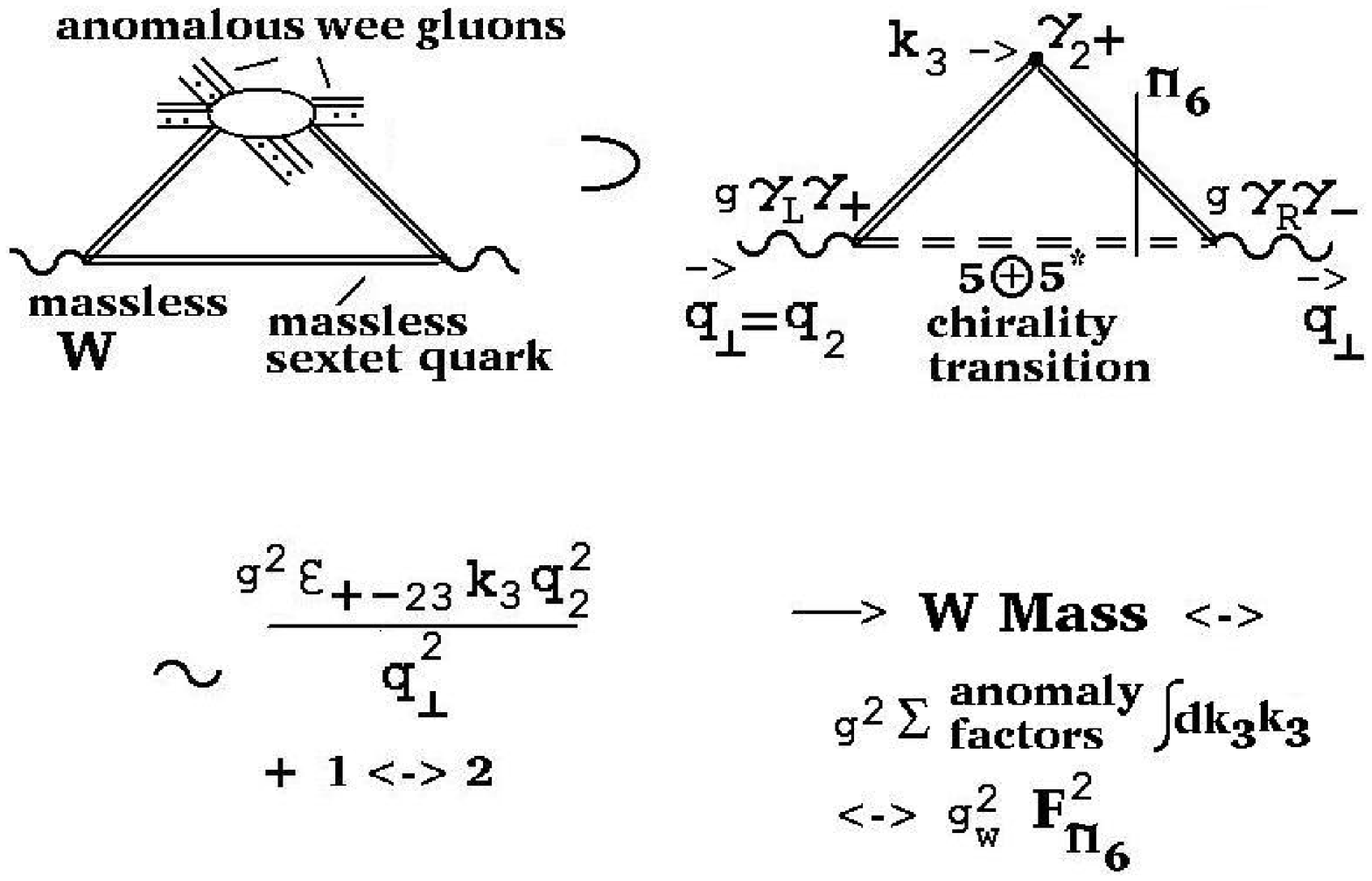}

Figure 4 Vector Boson Mass Generation. 
\end{center}
The chirality transition involved 
implies that the exchanged boson carries a ``right-handed'' fermion flavor quantum number that eliminates potential exponentiating infra-red divergent interactions. After the full SU(5) gauge symmetry is restored, and sextet couplings to the wee gauge bosons dominate, the fermion flavor becomes (dominantly) the right-handed sextet quark flavor symmetry and provides the physical SU(2)$_L$
symmetry. Because of the necessity for anomaly couplings, only states
containing (electroweak) SU(2) doublets couple to the SU(2)$_L$ interaction. 

\mainhead{6. THE QUD TOP QUARK AND EXOTICS}

According to the arguments in \cite{arwdm} and \cite{arw12} there are two straightforward ``Standard Model'' triplet quark generations in QUD, while the third generation
$$
(3,-\frac{1}{3}) ~\equiv ~[3,1,-\frac{1}{3}] {\large ~\in 5}, ~~~~
(3,\frac{2}{3}) ~\in [3,2,\frac{7}{6}] {\large ~\in 45^*}
\hspace{1in}  \{G3\}$$
is more unconventional. Correspondingly, amongst the antiquarks there are two almost identical generations which mix appropriately for them to be identified as belonging to the first two generations. They are triplets with respect to the electroweak SU(2) gauge symmetry but are singlets under the physical SU(2)$_L$. In contrast with the quarks, the third generation of triplet antiquarks
$$
(3^*,-\frac{2}{3}) \equiv [3^*,1,-\frac{2}{3}] {\large \bf ~\in  40},~~~~~ (3^*,\frac{1}{3})\equiv [3^*,1,\frac{1}{3}] {\large \bf ~\in  45^*}
\hspace{0.8in} \{AG3\}
$$
is actually conventional. The chosen identification of antiquarks
requires\cite{arwdm} that the physical b quark be a mixture of the charge
-1/3 quark in $\{G3\}$ and the charge -1/3 quarks in the other two generations. 

The higher charge sector of ``exotic'' triplet quarks is
\newline \parbox{1.2in}{$~~$ {\large exotics {\Huge \{} }}
\parbox{4in}{
$$
(3,\frac{5}{3}) ~\in [3,2,\frac{7}{6}] {\large \bf ~\in 45^*}, ~~~
(3,-\frac{4}{3}) ~\equiv ~[3,1,-\frac{4}{3}] {\large \bf ~\in 45^*}
$$
$$
(3^*,-\frac{5}{3}) \in [3^*,3,-\frac{2}{3}] {\large \bf ~\in  40}, ~~(3^*,\frac{4}{3}) \in [3^*,3,\frac{1}{3}] {\large \bf ~\in  45^*}
$$}
\newline After the initial restoration of  SU(2)$_C$ color the physical states,  consisting of chiral symmetry Goldstone boson anomaly poles, composed of massless quarks, will not include any exotics because this sector has no chiral symmetry.
Therefore, we expect that the exotic sector will 
either have no bound-states or will form only very massive states. 

From \{G3\}, we see that the charge 2/3 top quark forms an electroweak SU(2) doublet with one of the exotic quarks.
As a result, the left-handed top quark $t_L$ and the right-handed 
antiquark $\bar{t}_R$  will have a physical electroweak coupling to the exotic sector. The left-handed $\bar{t}_L$ and the right-handed $t_R$ will not have this coupling. Because of the presence of the sextet quark sector, the SU(2)$_L$
interaction will increase with energy (in the S-Matrix) above the sextet scale
and so, after it's introduction, 
potential low mass states containing a single $t_L$ quark will be destabilized, giving an effect similar to the top quark having a very large intrinsic mass. 

\mainhead{7. PARITY PROPERTIES OF THE {\LARGE \bf $\eta_3$} AND THE {\LARGE \bf $\eta_6$}}

The neutral pseudoscalar $\eta_t$ reggeon state discussed in \cite{arw12} is a combination of the $\eta_{tL} \equiv t_L\bar{t}_R$ and the $\eta_{tR} \equiv t_R\bar{t}_L$,
in both of which the top quark and antiquark have opposite helicities. The $\eta_t$ 
will be one of the  initial pseudoscalar, triplet sector, chiral symmetry Goldstone bosons that form bound-state anomaly poles when the SU(2)$_C$ color symmetry is restored. If we appropriately combine the $\eta_t$ with all the neutral $\eta$ mesons containing triplet quark/antiquark pairs
we obtain the $\eta_3$, in which the reggeon state will be a flavor singlet with respect to the color triplet quark sector. Initially, the $\eta_3$ is the massless axial U(1) Goldstone boson of the triplet sector while, similarly, the $\eta_6$ is the massless axial U(1) Goldstone boson of the sextet sector.

As discussed in Section 5, The ``sextet pions'' become the (dominant part of the) longitudinal component of the electroweak vector bosons and leave only the $\eta_6$ flavor singlet pseudoscalar. 
Consequently, if the sextet pions are compared with the Standard Model Higgs scalars that give the vector bosons their masses, then the $\eta_6$ compares directly with the left-over scalar, i.e. the ``Higgs'',
even though the sextet sector does not duplicate the role of the Higgs sector in providing general masses.

Both U(1) axial symmetries are broken by the 
transverse momentum cut-off in the anomaly vertices. 
Also, as flavor singlets, the $\eta_3$ 
and the $\eta_6$ both couple to the 
(daughter of) the pomeron as illustrated (after SU(2)$_C$ restoration) in Fig.~5.
\begin{center}
\epsfxsize=2.9in \epsffile{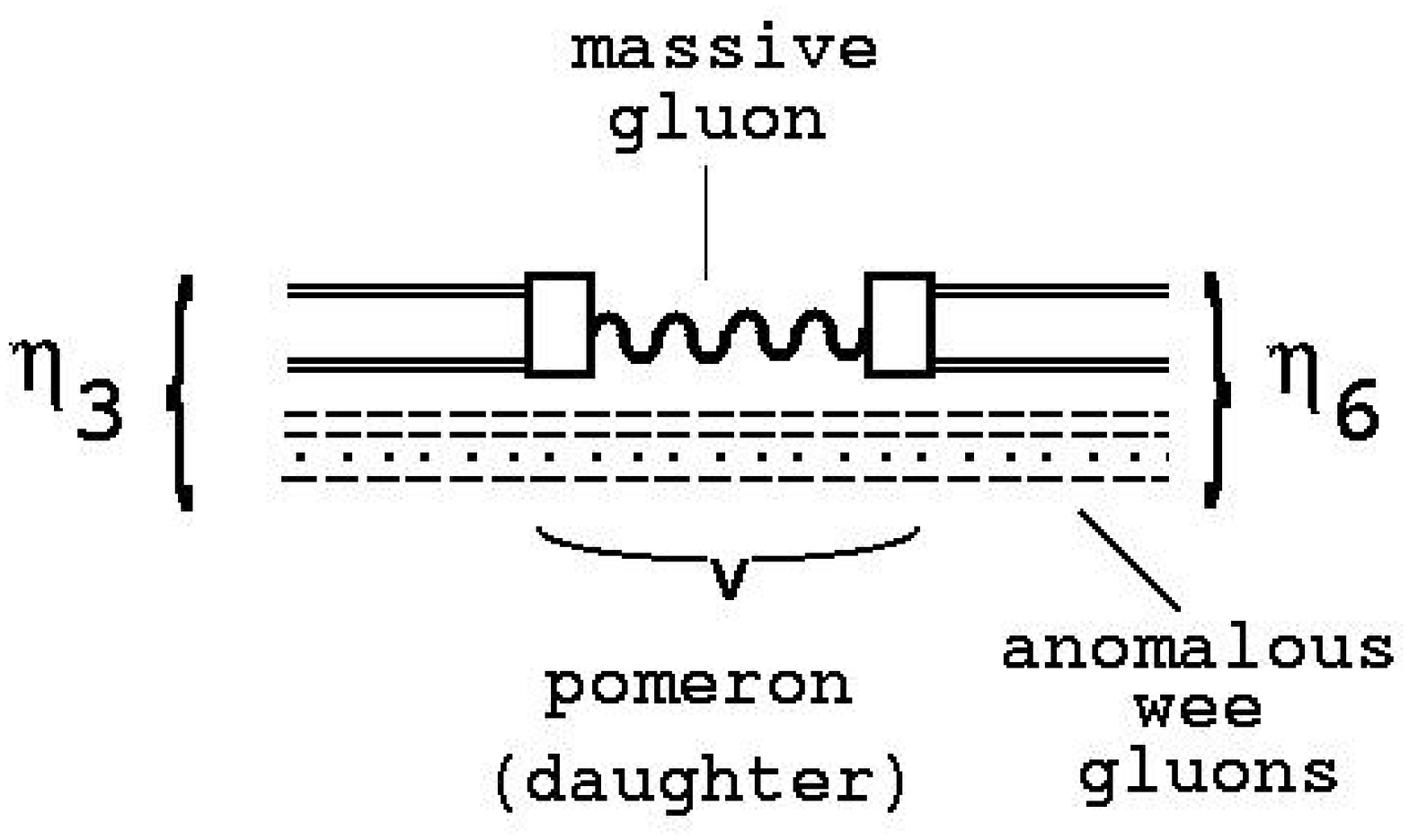}

Figure 5. Mixing of the {\large \bf $\eta_3$} and the {\large \bf $\eta_6$}
Reggeon States Via the Pomeron
\end{center}
This coupling ensures that both the triplet and the sextet U(1) symmetries
remain broken after the cut-off is removed and produces a mass for both Goldstone bosons. Because 
there is no bound-state (glueball) produced by the pomeron sector,
the resulting mixing produces no new resonances and so we
continue to call the two resonances, now mixed, the $\eta_6$ and the $\eta_3$. Without the electroweak interaction, both would be pseudoscalars.

The electroweak SU(2) gauge symmetry which produces the physical SU(2)$_L$ symmetry has a major effect on the $\eta_t$ components of the 
$\eta_3$ and $\eta_6$. The ``right-handed'' reggeon state $\eta_{t_R} \equiv t_R\bar{t}_L$ is a neutral state in which neither the quark or the antiquark have an SU(2) interaction. Contrastingly, in the ``left-handed'' reggeon state $\eta_{t_L} \equiv t_L\bar{t}_R$, both the quark and the antiquark do have an SU(2) interaction. Consequently, after the introduction of this interaction, the strong coupling to the exotic sector, via the SU(2)$_L$ interaction, will highly suppress the $\eta_{t_L}$ component of the $\eta_t$ and both the $\eta_3$ and the $\eta_6$ will lose their pseudoscalar parity. We expect this effect to be strongest in the lower mass $\eta_3$, which will continue to be
predominantly the $\eta_t$ and to have a mass closer to the triplet sector mass scale.

\mainhead{8. TREE-UNITARITY IN THE QUD S-MATRIX}

At first sight, the closeness of the electroweak reggeon diagrams to perturbation theory, as illustrated by the photon exchange amplitude in Fig.2, might 
suggest that we could directly study the tree-unitarity cancelation in the QUD S-Matrix. Unfortunately, since we can only describe bound-states and interactions as t-channel reggeon exchanges, we can not directly discuss the s-channel production processes involved. We could, perhaps, study the inter-relation of the analagous
t-channel processes shown in FIg.~6.
\begin{center}
\epsfxsize=4.5in \epsffile{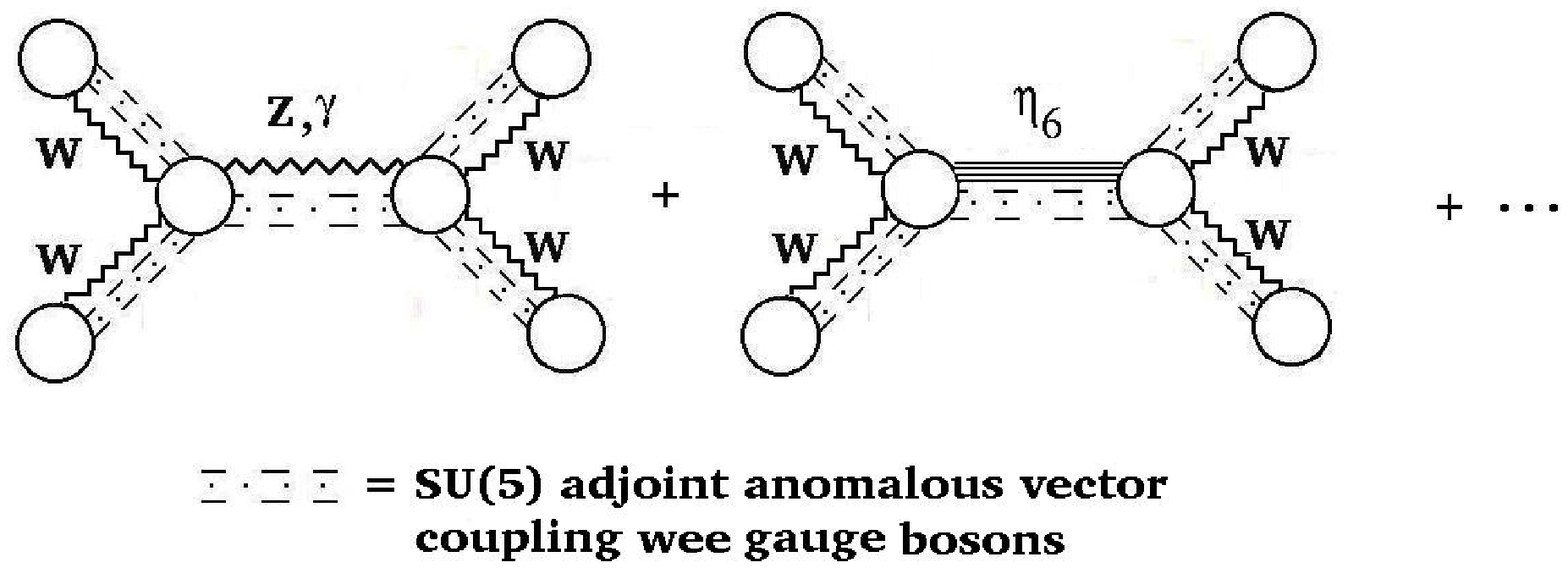}

Figure 6. Analagous t-channel Processes for Tree-Unitarity s-channel Processes.
\end{center}
Because of it's initial Goldstone boson status, the $\eta_6$ will share reggeon anomaly vertices with the sextet pions, as illustrated in Fig.~7, that would
appear in the processes illustrated in Fig.~6. 
After mixing, the $\eta_3$ will also have such vertices. Consequently, the interactions exist that could implement the tree-unitarity cancelation and, in principle at least, we could attempt to continue them to the s-channel. 
\begin{center}

\epsfxsize=4.5in \epsffile{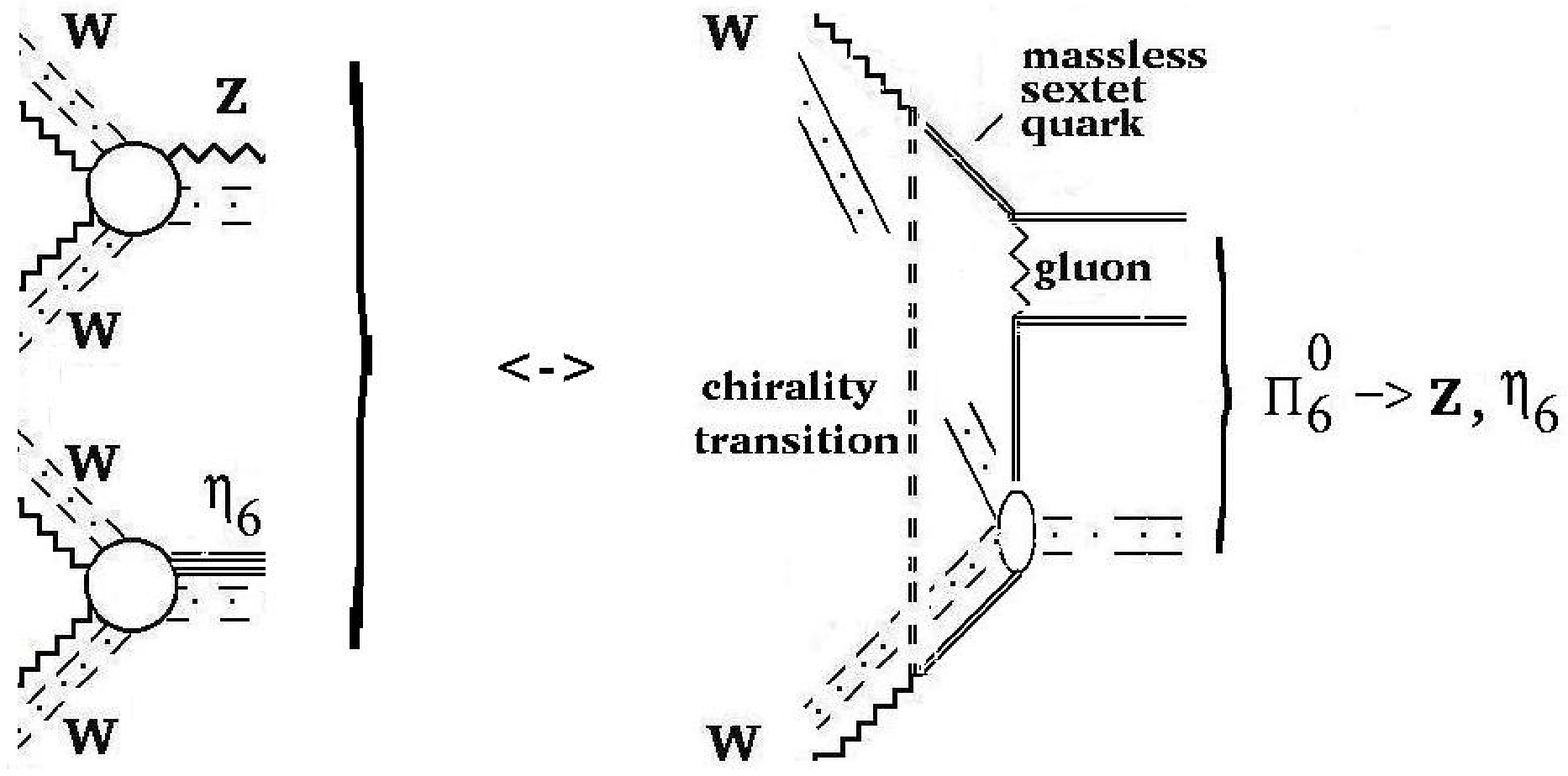}

Figure 7. Electroweak Anomaly Vertices Involving $W,Z$ and the $\pi_6, \eta_6$ Mesons
\end{center}

Unfortunately,
unknown details of the $\eta_6$ mixing with the $\eta_3$, added to the complicated anomaly vertex structure, surely implies that there is little hope of carrying out an explicit calculation, comparable with the Standard Model calculation\cite{lls,mt}, in the immediate future.
Nevertheless, as I have emphasized, if the QUD S-Matrix electroweak vector boson is a regge pole, as the reggeon diagram construction implies, then, as discussed in Section 3, the production of additional inter-related s-channel states must cancel to preserve the regge behavior. The only
states produced with appropriate sextet quark couplings will be the $\eta_6$ and the
$\eta_3$. Therefore, the generality of the regge behavior requirement is perhaps, the strongest argument for tree-unitarity in the QUD S-Matrix. 

\mainhead{9. NO LHC NIGHTMARE SCENARIO!}

At present, it could still be that the only outcome of the LHC experimental program 
is the existence of a ``Higgs-like'' resonance with properties that, within experimental
error, are consistent with it being the Standard Model Higgs boson. This would
leave the field in a tortured state of uncertainty as to how to go forward.
As I said in the Introduction, the theoretical inconsistency of the theory
would say there must be more, but there would be no indication as to what should be looked for. This is now regarded as the ``nightmare scenario''.
A common view\cite{woit} is that false expectations were raised because a 

{\it ``raft of heavily promoted speculative and unconvincing schemes for "Beyond Standard Model" physics all promised exciting new phenomena to be found at LHC-accessible energies''}

In most extensions of the Standard Model, supersymmetry 
has been a central ingredient and, because of it's overwhelming aesthetic and mathematical appeal, it  has been almost unimaginable, to a large part of the community, that it would not be discovered. From my perspective, supersymmetry has always seemed dangerously unphysical. Having been 
interested in complex dynamical solutions to physical multiparticle problems in which fermions (particularly via anomalies) and gauge bosons (without anomalies) clearly have wildly different dynamical roles, it has been
hard to believe that the real world complexity of the physical S-Matrix could emerge if any form of supersymmetry is imposed on a theory.

Five years ago {\it (before the LHC program had even begun)}
I said\cite{arw08} 

{\it ``The viewpoint that the more difficult dynamical problems 
in QCD can be put aside because they are not fundamental for going beyond the 
Standard Model, has clearly been a major factor in allowing the unlimited
speculation ... the 
freedom of invention associated with 
the guiding principle/paradigm that progress will come via inspired guesses for missing  
short-distance physics, combined with experimental 
verification via related rare processes, has not yet received any 
experimental confirmation and, most importantly, there is no historical precedent
for assuming that it will. ...the long distance regge region physics of QCD 
is not well understood
and ... it is by getting this physics} (multiparticle t-channel
unitarity) {\it  right that we are 
led directly to the underlying physics of the full Standard Model.''}

The conclusion from the arguments of the present paper is that there could be underlying physics that looks, at first, like the Standard Model, but is actually a much 
more subtle phenomenon. This possibility, at first sight, may not seem very different from those (of which there remain many) in which all the additional
``Beyond the Standard Model'' physics is pushed up to energies where it won't
be seen in the forseeable future. The difference is that, in the QUD S-Matrix, there is almost nothing beyond the physics of the Standard Model. There is only the sextet baryons (responsible for dark matter) and they should be produced at 
LHC energies, even if they are not detectable within the present framework. In my previous papers\cite{arw10}-\cite{arw12}, I have listed a number of experimental results  that provide evidence for the new physics but, arguably, do not yet make the case definitively. A major difficulty is distinguishing a characteristically short distance process from a kinematically fringe component of the new strong interaction provided by QUD. As I keep 
returning to, it is essential to look beyond the short-distance region 
currently explored at the LHC. It would indeed be tragic if dogmatic insistence 
that the new physics to be searched for is to be found at short-distance 
leads only to higher-energy, ever more intense, LHC running that is unable to see anything beyond the nightmare scenario. 

Since electroweak symmetry breaking is a QCD effect of the sextet 
sector within QUD, once high-enough energies are obtained (hopefully, the eventual LHC energy will be sufficient) this new strong interaction will surely require exploration at all momentum scales. A rich experimental program, which could 
include\cite{arwdm} large cross-sections for multiple electroweak vector bosons
spread across the rapidity axis (and even appearing in the double pomeron cross-section), very high multiplicities of associated 
soft hadrons, strong interaction production of dark matter sextet neutrons (neusons), amongst the new phenomena to be seen. 

It is hard to predict what the practical consequences would be of the discovery of a new, stable, very massive, strongly interacting, form of {\bf dark matter} that could produce enormously energetic matter-antimatter annihilation. It is not hard to imagine that some very dramatic possibilities would be included and that, surely, the idea that LHC discovery physics is irrelevant for real world practical problems would be completely negated!
Most likely, a window into this physics could be opened, if only briefly, by a series of lower luminosity runs as a prelude to the planned LHC upgrade.

\end{document}